\documentclass[twocolumn,tighten,trackchanges]{aastex63}
\usepackage{mathtools}
\usepackage{txfonts} 
\usepackage{hyperref}
\usepackage{color}

\turnoffeditone

\defcitealias{PaperI}{EHTC~I}
\defcitealias{PaperII}{EHTC~II}
\defcitealias{PaperIII}{EHTC~III}
\defcitealias{PaperIV}{EHTC~IV}
\defcitealias{PaperV}{EHTC~V}
\defcitealias{PaperVI}{EHTC~VI}
\defcitealias{PaperVII}{EHTC~VII}
\defcitealias{PaperVIII}{EHTC~VIII}
\defcitealias{Narayan_2021}{N21}
\defcitealias{Gelles_2021}{G21}
\defcitealias{Deglinnocenti_1985}{LD85}

\newcommand{\m}{M87*}
\newcommand{\s}{Sgr\,A*}
\begin{document}                                              

\title{Photon Orbit Signatures in Spectra of Black Hole Accretion Disks}
\shorttitle{}

\author[0000-0002-7179-3816]{Daniel~C.~M.~Palumbo}
\affiliation{Black Hole Initiative at Harvard University, 20 Garden Street, Cambridge, MA 02138, USA}
\affil{Center for Astrophysics $\vert$ Harvard \& Smithsonian, 60 Garden Street, Cambridge, MA 02138, USA}

\author[0000-0001-6952-2147]{George N.~Wong}
\email{gnwong@ias.edu}
\affiliation{School of Natural Sciences, Institute for Advanced Study, 1 Einstein Drive, Princeton, NJ 08540, USA}
\affiliation{Princeton Gravity Initiative, Princeton University, Princeton, New Jersey 08544, USA}

\author[0000-0001-5287-0452]{Angelo Ricarte}
\affiliation{Black Hole Initiative at Harvard University, 20 Garden Street, Cambridge, MA 02138, USA}
\affil{Center for Astrophysics $\vert$ Harvard \& Smithsonian, 60 Garden Street, Cambridge, MA 02138, USA}
\email{angelo.ricarte@cfa.harvard.edu}

\begin{abstract}

Light orbiting an accreting black hole may impact the disk or jet multiple times before escaping to the observer, at a variety of angles with respect to the local magnetic field. In this letter, we characterize the imprints of these long path lengths and disparate magnetic field impacts in synchrotron spectra of hot accretion disks, as the strongly lensed ``photon ring'' exhibits a higher synchrotron turnover frequency in each lensed sub-image. We apply tools of varying complexity: \edit1{first, we develop a minimal, unlensed one-zone model that isolates the first two sub-images of the accretion flow. By varying the magnetic field geometry encountered by each sub-image, we show that distinctive spectral signatures emerge in both total intensity and fractional linear polarization}. Second, we examine a semi-analytic radiatively inefficient accretion flow (RIAF) model, in which we find that there is generally a frequency at which the first indirect image outshines the direct image even in total flux density. Lastly, we demonstrate that even general relativistic magnetohydrodynamic (GRMHD) simulation snapshots show this spectral character. We find a typical correction to the unresolved spectrum of order $10\%$ near the turnover frequency \edit1{that grows with increasing viewing inclination}, growing to order unity at higher frequencies. We predict sensitive spectral studies of the cores of Messier 87* and Sagittarius A* at frequencies exceeding $300$ GHz to constrain the existence of the photon ring even without imaging, with prospects for photon ring detection even in other sources with unresolved shadows.

\end{abstract}

\section{Introduction}

Black holes are unique among commonplace astrophysical objects in that light itself can make full orbits around them. Recent decades have shown rich structure in bound photon orbits around spinning black holes, as each permitted orbital radius corresponds to a single photon angular momentum, while those photons which are perturbed and escape to infinity paint a picture of the spacetime itself in their morphology \citep[see, e.g.][]{Bardeen_1973, Teo_2003, Johannsen_2010}. 

A recent flurry of theoretical activity has given rise to renewed interest in observing these strongly lensed photons as they manifest in the ``photon ring,'' a ring of light expected in any image of a black hole for which the near-horizon region is optically thin \citep{Johnson_2020}. For emission geometries that have large differences in emissivity across the near-horizon region (such as accretion disks that grow dim near the poles), the photon ring takes on a ``wedding cake'' structure of stacked, exponentially demagnified lensed images of the accretion disk, with each tier of the cake corresponding to the photons that escape to infinity after increasing numbers of half-orbits around the black hole, as indexed by the integer $n$.

The first strongly lensed image, the $n=1$ sub-image, is a strong indicator for the presence of a black hole and a rich probe of the spacetime. Near-term very-long-baseline interferometry (VLBI) observations enabled by the next-generation Event Horizon Telescope \citep[ngEHT, see][]{Doeleman_2023} expansions of the EHT  are likely to be sufficient to detect the presence of the ring in Messier 87* (hereafter \m{}) and Sagittarius A* (hereafter \s{}), while the Black Hole Explorer (BHEX) mission is poised to precisely measure the shape and relative astrometry of the $n=1$ rings of these two sources, which are expected to serve as exquisite probes of the masses and spins of the black holes themselves \citep[][]{Johnson_2024, Lupsasca_2024}.

For virtually all other known supermassive black holes, detection of the first lensed sub-image has been a distant dream, limited by the angular size and brightness of targets beyond \m{} and \s{}. In order to demonstrate the existence of the photon ring in these sources, a booming signature in observations that do not spatially resolve the horizon would be necessary. For this, we turn to spectroscopy.

Black holes have a long history of rich and subtle spectral effects; the spectrum of Hawking radiation from the horizon carries quantum mechanical corrections that turn an otherwise black body gray \citep{Hawking_1975, Page_1976a, Page_1976b, Parikh_2000}.  In this letter, we instead look outside the horizon to the firmly astrophysical regime of hot accretion disks to show that the presence of photon orbits gives rise to a notable spectral signature in synchrotron-emitting plasmas, as the longer optical path lengths sampled by lensed sub-images see a synchrotron spectrum peaked at a higher frequency. \edit1{We first consider simple one-zone models for the accretion disk or jet of \m{} containing either thermal or power law-distributed electron energies. To account for contributions from a wide collection of emitting plasma regions, we then use ray-traced semi-analytic models of radiatively inefficient accretion flows (RIAFs) tuned to the more typical \s{} mass but varying in density. Finally, we return to \m{} in a full general relativistic magnetohydrodynamic (GRMHD) simulation, and find that the spectral imprint of the $n=1$ ring contributes at least a significant spectral flattening near the synchrotron turnover frequency and can in some cases create obvious spectral kinks at the peak frequency of the $n=1$ spectrum.}

\edit1{}

We begin with the analytic construction of the unlensed toy model in \autoref{sec:slab}. We move on to study the spectra of RIAFs and the GRMHD model in \autoref{sec:raytracing}. 
We conclude in \autoref{sec:conclusion}.

\section{A One-Zone, Two-Angle Photon Ring Model}
\label{sec:slab}

We begin by maximally simplifying the lensing and radiative transfer problem of relevance to the spectroscopy of the photon ring. We consider a stationary emitting cube of uniformly magnetized plasma with either thermally- or power law-distributed electron energies. We now review the applicable fully polarized radiative transfer problem in flat space so that we may study both total intensity and polarized signatures.

\begin{figure}
    \centering
    \includegraphics[width=0.5\textwidth]{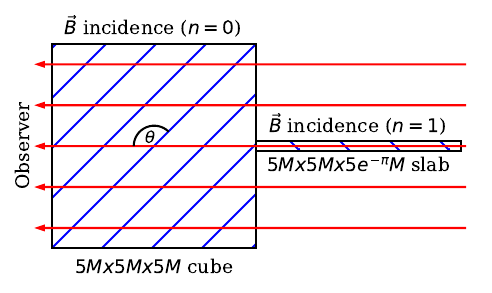}
    \caption{Cross-section of the one-zone, two-angle model for the spectroscopy of the $n=0$ and $n=1$ images. Blue lines indicate field lines, while red rays show the optical path. The same uniform plasma occupies both cells, but the two cells make different angles of incidence with the rays, and the reduced on-sky area of the $n=1$ image is represented by shrinking one dimension of the cube by $e^{-\pi}$. In this example for a $5M$ cube emitter, the $n=0$ magnetic field incidence angle is $135^\circ$, while the $n=1$ is $45^\circ$; rays passing through the $n=1$ region still experience radiative transfer through the $n=0$ region. Both regions are assumed to be at rest.}
    \label{fig:slab}
\end{figure}

\begin{figure*}[ht!]
    \centering
    \includegraphics[width=\textwidth]{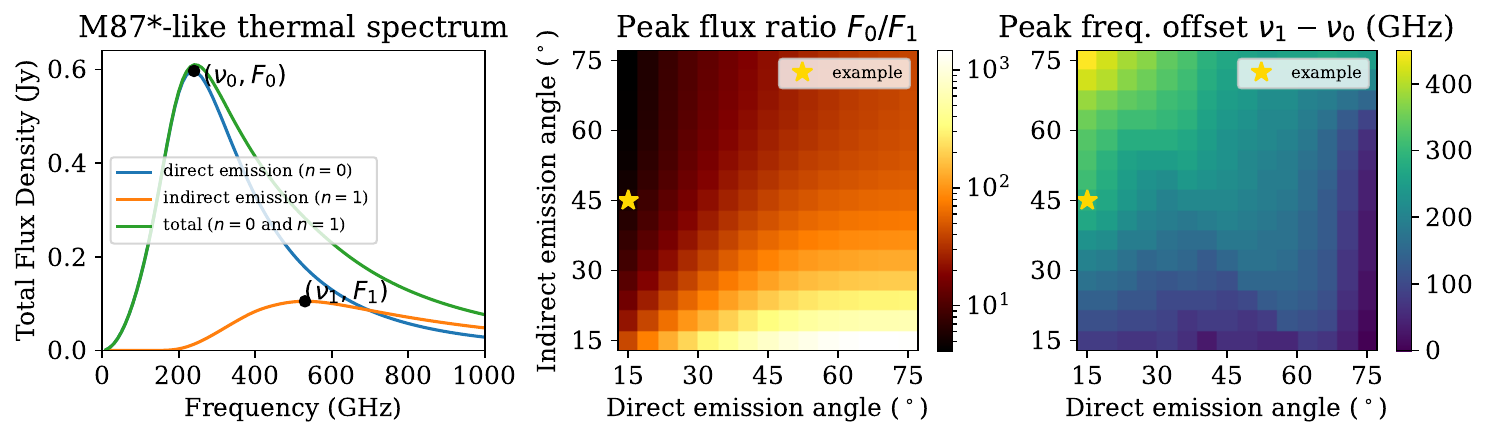}
    \includegraphics[width=\textwidth]{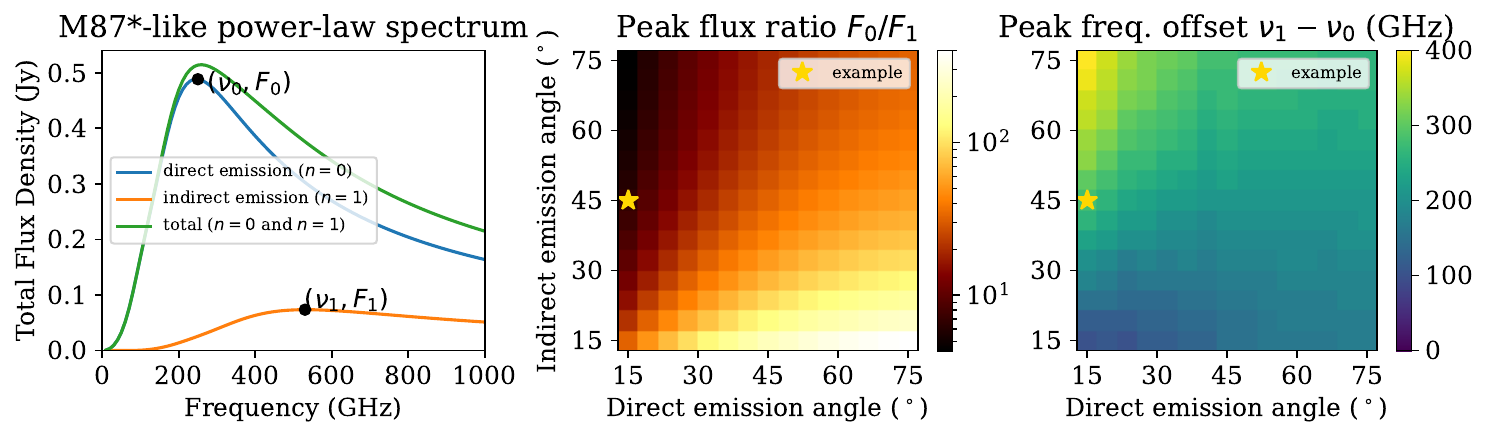}
    \caption{Spectroscopy of the photon ring as predicted by a one-zone model viewed from two angles, tuned to approximate the size and compact flux of the M87* accretion flow. Top row: thermal electron energy distribution with $\Theta_e = 10$ and $n_e=3\times10^4 {\rm cm}^{-3}$. Bottom row: power-law electron energy distribution with $p=3,\gamma_{\rm min}=100,\gamma_{\rm max}=10000$, and $n_e=5\times10^2 {\rm cm}^{-3}$. Both models have $B=30$~G. Left: an example spectrum from the model chosen to accentuate the effect, for which the $n=1$ image contributes a prominent spectral flattening beyond the turnover frequency. Middle: the range of ratios between the spectral peak of each sub-image as a function of the emission angles in each image. Right: same as the middle, but for the frequency offset between the peaks.}
    \label{fig:onezone_grid}
\end{figure*}

We consider the case of constant emission, absorption, and rotation coefficients in a uniformly magnetized region of synchrotron-emitting plasma. Following \citet{Deglinnocenti_1985}, hereafter \citetalias{Deglinnocenti_1985}, we begin with the radiative transfer equation paramterized in terms of the path length $s$ into the emitting material:
\begin{align}
    \frac{d\bf{I}}{ds} & = {\bf j}(s)- {\bf K}(s) {\bf I}(s),\\
    \rightarrow \frac{d\bf{I}}{ds} & = {\bf j}- {\bf K} {\bf I} (s),
\end{align}
where 
\begin{align}
    {\bf j} &= \left(j_I, j_Q, j_U, j_V \right)^\dagger,\\
    {\bf I} &= \left(I, Q, U, V\right)^\dagger.
\end{align}

Here, ${\bf K}$ is the transfer matrix:
\begin{align}
{\bf K} & \equiv
\begin{pmatrix}
    \alpha_I & \alpha_Q & \alpha_U & \alpha_V\\
    \alpha_Q & \alpha_I & \rho_V & -\rho_U\\
    \alpha_U & - \rho_V & \alpha_I & \rho_Q\\
    \alpha_V & \rho_U & -\rho_Q & \alpha_I,
\end{pmatrix}
\end{align}
where each $\alpha$ is the absorption coefficient and each $\rho$ is the rotation coefficient associated with each Stokes parameter of polarization. The source function vector $\mathbf{S} \equiv (j_I / \alpha_I, j_Q/\alpha_Q, j_U/\alpha_U, j_V/\alpha_V)^\dagger$ and the transfer operator ${\bf O}(s) = \exp (-{\bf K} s)$ for the case where $\bf{K}$ is constant in space. The solution to the radiative transfer equation is then
\begin{align}
    {\bf O}(x) &= \exp(- {\bf K} x),\\
    {\bf I}(s) &= \int_{0}^{s} {\bf O}(s'){\bf j} {\rm d}s',\\
    &= \left[\int_{0}^{s}{\bf O}(s')ds'\right]{\bf j},\nonumber\\
    &\equiv {\bf T}(s) {\bf j}. \label{eq:rad}
\end{align}
\citetalias{Deglinnocenti_1985} found a closed form for $\bf O$ such that the integral above can be performed analytically. \citet{IPOLE_2018} presented this integral; we choose another equivalent organization of terms with four sub-integrals, which we provide below:
\begin{align}
    T_1 &= \int_0^s {\rm d}x e^{- \alpha_I x} \cosh{\Lambda_1 x} \nonumber \\
    &= \frac{1}{2}\left(\frac{1-e^{ -( \alpha_I - \Lambda_1)s}}{\alpha_I-\Lambda_1} + \frac{1-e^{- ( \alpha_I+\Lambda_1)s}}{\alpha_I+\Lambda_1}\right)\\
    T_2 &= \int_0^s {\rm d}x e^{- \alpha_I x} \sin{\Lambda_2 x} \nonumber \\
    &= e^{-\alpha_I s}\frac{\Lambda_2 \left[e^{\alpha_I s}-\cos(\Lambda_2 s)\right] - \alpha_I \sin(\Lambda_2 s) }{\alpha_I^2 + \Lambda_2^2}\\
    T_3 &= \int_0^s {\rm d}x e^{- \alpha_I x} \sinh{\Lambda_1 x} \nonumber \\
    &= \frac{1}{2}\left(\frac{1-e^{ -( \alpha_I-\Lambda_1)s}}{\alpha_I-\Lambda_1} - \frac{1-e^{- ( \Lambda_1+\alpha_I)s}}{\alpha_I+\Lambda_1}\right)\\
    T_4 &= \int_0^s {\rm d}x e^{- \alpha_I x} \cos{\Lambda_2 x} \nonumber \\
    &= e^{-\alpha_I s}\frac{\alpha_I \left[e^{\alpha_I s}-\cos(\Lambda_2 s)\right] + \Lambda_2 \sin(\Lambda_2 s) }{\alpha_I^2 + \Lambda_2^2}
\end{align}
The resulting integrated transfer operator, which we call ${\bf T} \equiv\int_{-\infty}^{s}{\bf O}(s')ds'$, is given by
\begin{align}
    {\bf T} &= \frac{1}{2}(T_1+T_4) {\bf M_1} - T_2 {\bf M_2} - T_3 {\bf M_3} + \frac{1}{2} (T_1-T_4){\bf M_4}.
\end{align}
Here, each matrix ${\bf M}$ is as given in \citetalias{Deglinnocenti_1985}. Thus, using Equation \ref{eq:rad}, all that is required for a prediction of the Stokes intensities is the full set of transfer coefficients, for which we turn to the analytic fitting formulae in \citet{Marszewski_2021}. These coefficients build on work by \citet{Leung_2011}, \citet{Pandya_2016}, \citet{Dexter_2016}, and many others.

\begin{figure}[t!]
    \centering
    \includegraphics[width=\linewidth]{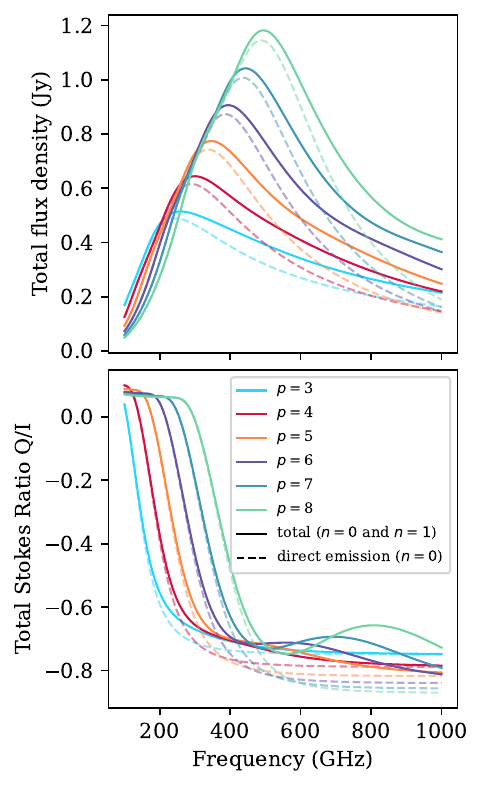}
    \caption{Survey of power law index $p$ for \edit1{the direct and indirect image magnetic field incidence angles} $\theta_{B,0}=15^\circ$ and $\theta_{B,1}=45^\circ$ as shown in \autoref{fig:onezone_grid}. Steeper power laws increase the relative contribution of the photon ring, leading to substructure in the spectrum of the polarization fraction.}
    \label{fig:powerlaws}
\end{figure}

\begin{figure*}[t!]
    \centering
    \includegraphics[width=\textwidth]{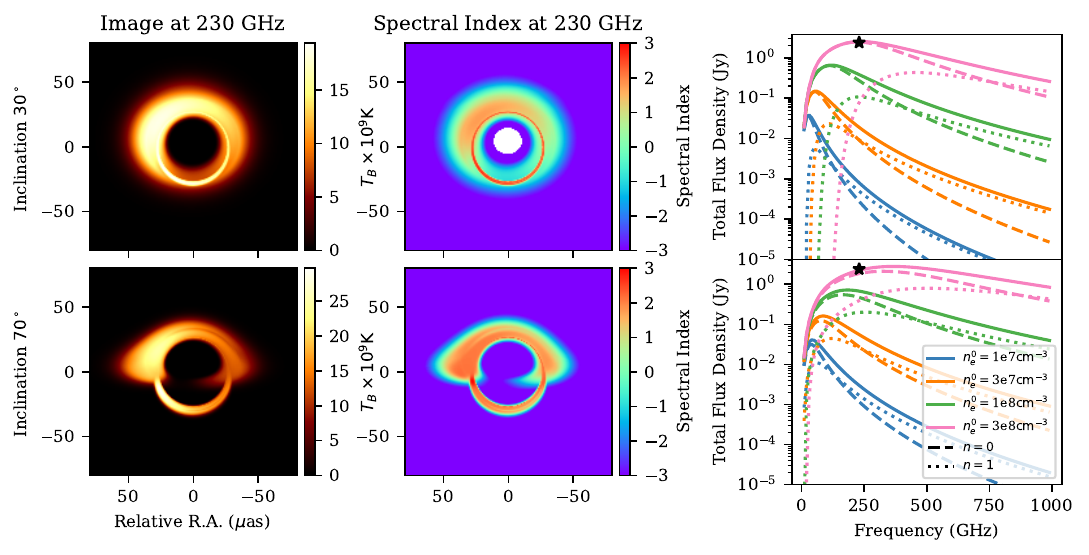}
    \caption{Dependence of the photon ring spectral feature on inclination and electron density in a RIAF model of \s{}. Left: 230 GHz images of the same model at $30^\circ$ and $70^\circ$ inclination, with a fixed peak number density of $3\times10^8{\rm cm}^{-3}$. Middle: spectral index maps at 230-235 GHz at each inclination. Right: spectra of the decomposed model at each inclination, varying the electron number density. The black star indicates the 230 GHz, 2.4 Jy value at the density used in the left figures.}
    \label{fig:RIAF_summary}
\end{figure*}

Neglecting for the moment the details of electron thermodynamics and recalling that we hold the plasma to be at rest with respect to the viewer, the resulting one-zone cube model predicts received Stokes parameters as a function of the following parameters:
\begin{enumerate}
    \item $\theta_B$, the inclination of the line of sight to the magnetic field,
    \item $s$, the side length of the emitting cube,
    \item $|B|$, the magnitude in Gauss of the magnetic field threading the plasma,
    \item $n_e$, the number density of electrons in the plasma, in units of ${\rm cm}^{-3}$.
\end{enumerate}
We include the $n=1$ image by placing another identical copy of the same one-zone emitter \textit{behind} the first emitting cube, but shrunk along one axis by $e^{-\pi}$, in accordance with the limiting geometric demagnification of subsequent photon ring images \citep{Johnson_2020}. This slab's emission is transferred through the foreground cube on the way to the observer (albeit with the emissivities in the foreground cube set to zero), as diagrammed in \autoref{fig:slab} for a cube of side length $5M$, where we have used gravitational units $M\equiv GM_{\rm BH}/c^2$, with gravitational constant $G$, black hole mass $M_{\rm BH}$, and speed of light $c$. Distinct emission geometries and lensing properties are approximated entirely through the changing of the magnetic field pitch angles encountered by a ray as it passes through the two emitting regions.

In addition to these general parameters, the electron thermodynamical prescription also requires specification. We consider electron energies distributed either thermally according to a dimensionless electron temperature $\Theta_e = k_B T / m_e c^2$ (where $m_e$ is the electron mass, $c$ is the speed of light, $k_B$ is the Boltzmann constant, and $T$ is the temperature in Kelvin), or a power law in Lorentz factor $\gamma$ between some $\gamma_{\rm min}$ and $\gamma_{\rm max}$ with a power law index $p$.

\autoref{fig:onezone_grid} shows the spectral properties of this one-zone model for an \m{}-like emitting cube, taking $M=10^{15}$cm and  $|B|=30{\rm G}$ and varying $n_e$ to approximate the \m{} compact flux, corresponding to the less dense but more magnetized end of one-zone models consistent with the data analysis in \citet{PaperV} and \citet{PaperVIII}. We consider both thermal electrons with $\Theta_e=10$ and power-law electrons with index $p=3$, $\gamma_{\rm min}=100$, and $\gamma_{\rm max}=10000$. We compute the spectra for the $n=0$, $n=1$, and total images, and compute how their peak fluxes and frequencies vary as a function of the angle of incidence of rays on the magnetic fields in each region, \edit1{explored in the right two columns}. As expected, as the rays become perpendicular to the magnetic field, the synchrotron emissivity (and absorptivity) increases, causing the largest fractional contributions from the $n=1$ when the direct emission angle is shallow while the indirect emission angle is steep. We observe that the sharper fall-off in high energy electrons in the thermal distribution allows the $n=1$ spectrum to more readily dominate at high frequencies when permitted by the distinct magnetic field inclinations.

In \autoref{fig:powerlaws}, we survey the total flux density and linear polarization fraction (expressed as the signed ratio of Stokes $Q$ and $I$) in the example power law electron model shown in \edit1{the bottom left panel of} \autoref{fig:onezone_grid}, which has $\theta_{B,0}=15^\circ$ and $\theta_{B,1}=45^\circ$. In the absence of the $n=1$ image, we would expect a transition between optically thick and thin fractional polarizations given by \citep{R&L}:
\begin{align}
    \Pi_{\rm thick} &= \frac{3}{6p+13},\\
    \Pi_{\rm thin} &= \frac{-(p+1)}{(p+\frac{7}{3})}.
\end{align}
We find that steeper power laws have a larger relative contribution at high frequencies, with a larger corresponding difference in the polarization fraction at intermediate frequencies at which the $n=1$ image becomes optically thin. For steeper power laws or magnetic field impact angles that more steeply favor the $n=1$ image, the $n=1$ polarization signal causes a kink towards lower polarization fraction before turning back towards the high-frequency limit.

\section{RIAFs and GRMHD}
\label{sec:raytracing}

To incorporate some of the subtleties of a near-horizon accretion disk to our spectral analyses, we utilize radiatively inefficient accretion flow (RIAF) models. We use a model described in \citet{Wong_2025} and patterned after the RIAF used to fit \s{} observations in \citet{Pu_2018}. The new model differs primarily in its prescriptions for fluid velocity and magnetic field structure. The fluid velocity is taken to follow the form given in \citet{AART_2023}, which more closely matches general relativistic magnetohydrodynamic (GRMHD) simulations. 
We also use the monopolar magnetic field configuration described in \citet{Chael_2023}, which obeys regularity conditions at the event horizon and has been shown to qualitatively reproduce the magnetic field structure in the hot magnetosphere above the accretion disk.
However, our electron number density ($n_e$) and electron temperature ($T_e$) still vary according to the same power laws in the radius, $r$:
\begin{align}
    n_e &= n_e^0 r^{-\alpha}e^{-\beta},\\
    T_e &= T_e^0 r^{-\gamma}.
\end{align}
We vary $n_e^0$ and take $T_e^0=3\times10^{11}$ K, $\alpha=1.1$, and $\gamma=0.84$ throughout to maximally align with the models in \citet{Pu_2018}. We use a fractional disk height $H$ of 0.1, which sets $\beta$ through its dependence on the disk thickness $\sigma$:
\begin{align}
    \sigma &= H r \sin \theta,\\
    \beta &= \frac{(r \cos\theta)^2}{\sigma^2}.
\end{align}
The electron energy distribution is everywhere thermal.

In \autoref{fig:RIAF_summary} we show 230 GHz images and spectral index maps for \s{} RIAFs with $n_e^0=3\times10^8{\rm cm}^{-3}$, as well as spectra from 1 GHz to 1 THz for RIAF models of varying electron number density. For our example models, which approximately match the compact flux density of \s{}, the model is near the synchrotron turnover frequency, meaning that optical depths in the image are near unity. As a result, the $n=1$ image peeks through the direct image and is occluded on the far side of the black hole, \edit1{apparent in both images in the left column}. At this transitional frequency, the spectral index maps show the $n=0$ image beginning to fall in brightness, while the $n=1$ image has a strongly positive spectral index.

\begin{figure*}[t!]
    \centering
    \includegraphics[width=\textwidth]{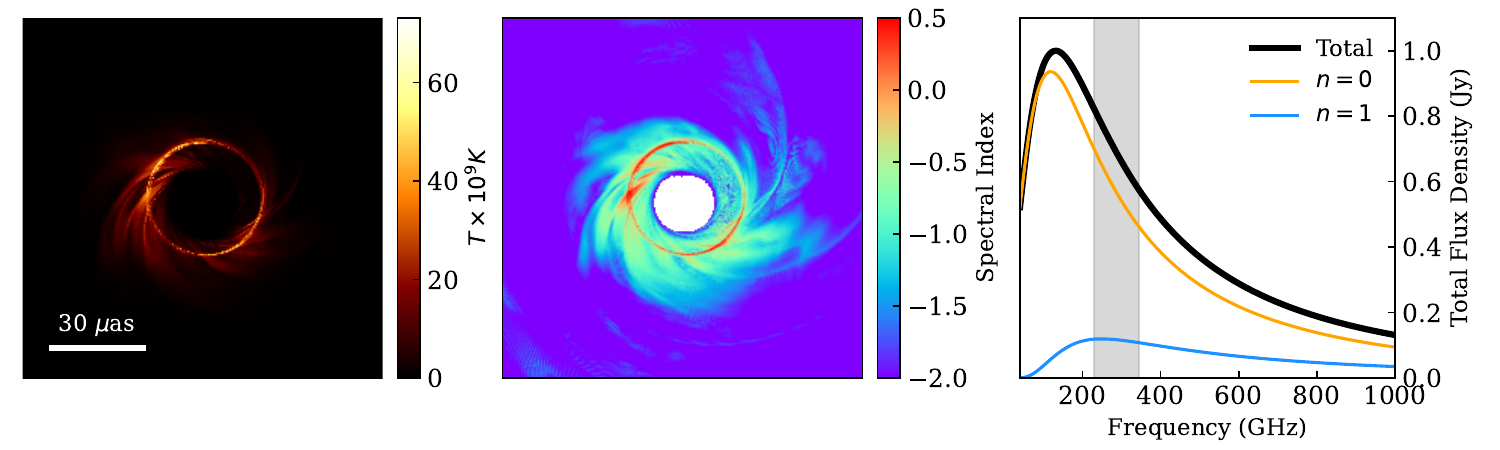}
    \caption{Spectral feature in ray-traced GRMHD, corresponding to a MAD $a=0.5$ $R_\mathrm{high}=160$ model of M87*.  Left:  total intensity map. Center: spectral index map at 230 GHz revealing the elevated spectral index in the photon ring.  Right: instantaneous spectrum, where the grey vertical band demarcates 230-345 GHz, and a circle marks the maximum of each contribution.  }
    \label{fig:grmhd}
\end{figure*}

We observe from the spectra in the right column of \autoref{fig:RIAF_summary} that \edit1{adding the $n=1$ image increases the total flux density at the spectral peak by $3-5\%$ at $30^\circ$ inclination, and $29-33\%$ at $70^\circ$ inclination. We might expect that steeper inclinations accentuate the spectral features of the photon ring due to the higher overall optical depths when the viewer lies near the plane of the disk.} Due in part to the strictly thermal electron energy distribution, we find that there always exists a frequency at which the $n=1$ image begins to outshine the $n=0$ image; \edit1{in the \s{}-like example with $n_e^0=3\times10^8{\rm cm}^{-3}$, this tipping point is at nearly 1 THz.} Empirically, this frequency tends to be $\sim3$ times the peak frequency of the complete spectrum. On the high frequency end of each spectrum, the contribution from the $n=1$ image can be viewed as an anomalous over-brightness that would otherwise require additional emitting material to ``lift'' the $n=0$ emission to the correct full spectrum.

In \autoref{fig:grmhd}, we plot the SED and a representative spectral index map of a GRMHD model of \m{}.  The simulation originates from \citet{Narayan_2022}, and is a MAD with $a=0.5$ and $R_\mathrm{high}=160$ \citep[as defined in][]{Mosci_2016}. Ray-tracing is performed following the Patoka pipeline \citep{Wong_2022}, with more specific details for these simulations available in \citet{Ricarte_2022} and \citet{Qiu_2023}.  A purely thermal electron distribution function is modeled. \edit1{We consider a single snapshot, unlike the time-averaged RIAF analysis. Over time, the accretion rate and magnetic field geometries may vary quickly about a stable mean (as is often observed in \s{}), or could drift, corresponding to variation across \autoref{fig:onezone_grid} or up or down in \autoref{fig:RIAF_summary}. However, an instantaneous GRMHD model provides the most realistic prediction for snapshot spectral measurements of the core of \m{}.}

As is expected for the significantly less dense plasma around \m{}, the model grows optically thin at a lower frequency than the putative \s{} RIAF examined earlier. As a result, the $n=1$ image has its spectral peak near the Event Horizon Telescope observational bands of 230-345 GHz. However, qualitatively, the high-frequency behavior of the spectrum is the same, with the relative contribution \edit1{to the total flux density} from the $n=1$ growing with frequency. \edit1{As in the RIAF model, the photon ring is apparent in the spectral index map due to its higher spectral index, in this case near zero across the ring. Again similarly to the RIAF model}, due to the inherently ``multi-zone'' nature of an accretion flow with varying plasma properties in the bulk, there is no sharp spectral kink as might be expected from the most favorable magnetic field geometries in a one-zone model.



\section{Conclusion}
\label{sec:conclusion}

We have identified the spectral imprint of orbiting light in the synchrotron-emitting accretion disk of a black hole. Due to the larger optical path lengths within the photon ring, more strongly lensed light corresponds to a spectrum peaked at higher frequencies. We have used a simplified one-zone model for a stationary emitting plasma to show that the angle of incidence with the magnetic field  in each sub-image can tune the relative dominance of the $n=0$ and $n=1$ image, and found that thermal electron energy distributions tend to show a more marked transition between subimages due to their more sharply falling spectra. Using a RIAF model, we showed that even in the presence of more realistic distributions of velocity and magnetic fields near the horizon, the photon ring smooths the spectrum near its peak, and eventually dominates the total received emission at higher frequencies at which both the $n=0$ and $n=1$ images are optically thin. We then confirmed the growing contribution of the $n=1$ image and its corresponding spectral flattening in an example GRMHD image.

Our results suggest that the existence of photon orbits should have a measurable imprint on the unresolved spectra of the cores of active galactic nuclei. As our analysis of the electron number density in the RIAF indicated, black holes that are accreting very little will be uniformly optically thinner, allowing unresolved hunts for the photon ring even at lower frequencies, while for \m{}, \s{}, and more highly accreting sources, the spectral imprints of the photon ring are expected in the strictly sub-mm regime. 

While ``unresolved'' photon signatures are exciting, we expect that several confounding effects that could mimic the spectral imprint of the photon ring would be mitigated by high spatial resolution. Firstly and most generally, for accretion disks and observing frequencies for which the photon ring does not create a sharp spectral knee and instead a gentle flattening, any integrative effect that sums contributions from many plasma regions would also have a net-flattening effect. Secondly and more specifically, synchrotron-emitting regions can produce their own spectral breaks that could reasonably imitate the presence of the $n=1$ ring, such as a cooling break \citep[see, e.g.][]{Sari_1996}. As a result, we anticipate that the most useful place to look for spectral photon ring signatures beyond \m{} and \s{} is in marginally resolved cores for which the downstream jet thermodynamics are well-understood (that is, which have extensive spectral index mapping that will disentangle cooling breaks as a function of distance along the jet). \edit1{We note that this requires less resolving power than the black hole shadow itself; as shown in \citet{PaperIV}, the interferometric baselines at which the shadow is able to be imaged exceed those dominated by the innermost jet by an order of magnitude in fringe spacing. For sources that are highly accreting, the near-THz compact flux may be sufficient to be detectable by ALMA even in bands 9 or 10, permitting photon ring searches in the array's most wide (and therefore highly resolving) configurations \citep[see, e.g.][]{ALMAmanual}.}

As discussed in \citet{Ricarte_2023} and shown again in \autoref{fig:RIAF_summary} and \autoref{fig:grmhd}, resolved spectral index maps of the inner accretion flow at frequencies just beyond the turnover for the direct image reveal a strikingly positive spectral index in the $n=1$ image. As sub-mm interferometry pushes to lower wavelengths and longer baselines, as well as simultaneous multi-wavelength observation \edit1{as in the case of the ngEHT and BHEX}, astrometrically aligned images in the 220-360 GHz regime will become commonplace. Even with limited resolution, we anticipate spectral flattening near the photon ring from beam mixing of the $n=0$ and $n=1$ image, with an eventual turnover at higher frequencies to a negative $n=1$ spectral index. \edit1{This beam flattening of the spectrum will decrease significantly at the $\sim6\mu$as nominal resolution of BHEX at 320 GHz on \m{}, compared to the $\sim14\mu$as resolution of the ngEHT at 345 GHz.} We predict that these multiwavelength observations will be highly constraining of not just the presence or absence of the photon ring, but also of the detailed plasma properties as elucidated by the distinct angles of incidence with the plasma sampled by highly bent rays.

\acknowledgments{We thank Dominic Pesce, Ramesh Narayan, and Michael Johnson for several helpful discussions. We are also grateful to our journal referee for their thoughtful comments on the manuscript. We acknowledge financial support from the National Science Foundation (AST-2307887). This work was supported by the Black Hole Initiative, which is funded by grants from the John Templeton Foundation (Grant 62286) and the Gordon and Betty Moore Foundation (Grant GBMF-8273) - although the opinions expressed in this work are those of the author(s) and do not necessarily reflect the views of these Foundations. G.N.W.\,was supported by the Princeton Gravity Initiative and by the Taplin Fellowship.}

\bibliography{main}

@ARTICLE{Ricarte_2023,
       author = {{Ricarte}, Angelo and {Gammie}, Charles and {Narayan}, Ramesh and {Prather}, Ben S.},
        title = "{Probing plasma physics with spectral index maps of accreting black holes on event horizon scales}",
      journal = {\mnras},
         year = 2023,
        month = mar,
       volume = {519},
       number = {3},
        pages = {4203-4220},
          doi = {10.1093/mnras/stac3796},
archivePrefix = {arXiv},
       eprint = {2202.02408},
 primaryClass = {astro-ph.HE},
       adsurl = {https://ui.adsabs.harvard.edu/abs/2023MNRAS.519.4203R}
}

@ARTICLE{Hawking_1975,
       author = {{Hawking}, S.~W.},
        title = "{Particle creation by black holes}",
      journal = {Communications in Mathematical Physics},
         year = 1975,
        month = aug,
       volume = {43},
       number = {3},
        pages = {199-220},
          doi = {10.1007/BF02345020},
       adsurl = {https://ui.adsabs.harvard.edu/abs/1975CMaPh..43..199H}
}

@ARTICLE{Page_1976a,
       author = {{Page}, Don N.},
        title = "{Particle emission rates from a black hole: Massless particles from an uncharged, nonrotating hole}",
      journal = {\prd},
         year = 1976,
        month = jan,
       volume = {13},
       number = {2},
        pages = {198-206},
          doi = {10.1103/PhysRevD.13.198},
       adsurl = {https://ui.adsabs.harvard.edu/abs/1976PhRvD..13..198P}
}

@MISC{ALMAmanual,
       author = {{Remijan}, A. and {Biggs}, A. and {Cortes}, P.~A. and {Dent}, B. and {Di Franceso}, J. and {Fomalont}, E. and {Hales}, A. and {Kameno}, S. and {Mason}, B. and {Philips}, N. and {Saini}, K. and {Vila Vilaro}, B. and {Villard}, E.},
        title = "{ALMA Technical Handbook,ALMA Doc. 7.3, ver. 1.1}",
 howpublished = {2019, ALMA Technical Handbook,ALMA Doc. 7.3, ver. 1.1ISBN 978-3-923524-66-2},
         year = 2019,
        month = jun,
          doi = {10.5281/zenodo.4511522},
       adsurl = {https://ui.adsabs.harvard.edu/abs/2019athb.rept.....R}
}

@ARTICLE{Page_1976b,
       author = {{Page}, D.~N.},
        title = "{Particle emission rates from a black hole. II. Massless particles from a rotating hole}",
      journal = {\prd},
         year = 1976,
        month = dec,
       volume = {14},
       number = {12},
        pages = {3260-3273},
          doi = {10.1103/PhysRevD.14.3260},
       adsurl = {https://ui.adsabs.harvard.edu/abs/1976PhRvD..14.3260P}
}

@ARTICLE{Parikh_2000,
       author = {{Parikh}, Maulik K. and {Wilczek}, Frank},
        title = "{Hawking Radiation As Tunneling}",
      journal = {\prl},
         year = 2000,
        month = dec,
       volume = {85},
       number = {24},
        pages = {5042-5045},
          doi = {10.1103/PhysRevLett.85.5042},
archivePrefix = {arXiv},
       eprint = {hep-th/9907001},
 primaryClass = {hep-th},
       adsurl = {https://ui.adsabs.harvard.edu/abs/2000PhRvL..85.5042P}
}

@ARTICLE{Sari_1996,
       author = {{Sari}, Re'em and {Narayan}, Ramesh and {Piran}, Tsvi},
        title = "{Cooling Timescales and Temporal Structure of Gamma-Ray Bursts}",
      journal = {\apj},
         year = 1996,
        month = dec,
       volume = {473},
        pages = {204},
          doi = {10.1086/178136},
archivePrefix = {arXiv},
       eprint = {astro-ph/9605005},
 primaryClass = {astro-ph},
       adsurl = {https://ui.adsabs.harvard.edu/abs/1996ApJ...473..204S}
}

@ARTICLE{Teo_2003,
       author = {{Teo}, Edward},
        title = "{Spherical Photon Orbits Around a Kerr Black Hole}",
      journal = {General Relativity and Gravitation},
         year = 2003,
        month = nov,
       volume = {35},
       number = {11},
        pages = {1909-1926},
          doi = {10.1023/A:1026286607562},
       adsurl = {https://ui.adsabs.harvard.edu/abs/2003GReGr..35.1909T}
}

@ARTICLE{AART_2023,
       author = {{C{\'a}rdenas-Avenda{\~n}o}, Alejandro and {Lupsasca}, Alexandru and {Zhu}, Hengrui},
        title = "{Adaptive analytical ray tracing of black hole photon rings}",
      journal = {\prd},
         year = 2023,
        month = feb,
       volume = {107},
       number = {4},
          eid = {043030},
        pages = {043030},
          doi = {10.1103/PhysRevD.107.043030},
archivePrefix = {arXiv},
       eprint = {2211.07469},
 primaryClass = {gr-qc},
       adsurl = {https://ui.adsabs.harvard.edu/abs/2023PhRvD.107d3030C}
}

@ARTICLE{Chael_2023,
       author = {{Chael}, Andrew and {Lupsasca}, Alexandru and {Wong}, George N. and {Quataert}, Eliot},
        title = "{Black Hole Polarimetry I. A Signature of Electromagnetic Energy Extraction}",
      journal = {\apj},
         year = 2023,
        month = nov,
       volume = {958},
       number = {1},
          eid = {65},
        pages = {65},
          doi = {10.3847/1538-4357/acf92d},
archivePrefix = {arXiv},
       eprint = {2307.06372},
 primaryClass = {astro-ph.HE},
       adsurl = {https://ui.adsabs.harvard.edu/abs/2023ApJ...958...65C}
}

@ARTICLE{Ricarte_2022,
       author = {{Ricarte}, Angelo and {Palumbo}, Daniel C.~M. and {Narayan}, Ramesh and {Roelofs}, Freek and {Emami}, Razieh},
        title = "{Observational Signatures of Frame Dragging in Strong Gravity}",
      journal = {\apjl},
         year = 2022,
        month = dec,
       volume = {941},
       number = {1},
          eid = {L12},
        pages = {L12},
          doi = {10.3847/2041-8213/aca087},
archivePrefix = {arXiv},
       eprint = {2211.01810},
 primaryClass = {gr-qc},
       adsurl = {https://ui.adsabs.harvard.edu/abs/2022ApJ...941L..12R}
}

@ARTICLE{Deglinnocenti_1985,
       author = {{Landi Degl'Innocenti}, E. and {Landi Degl'Innocenti}, M.},
        title = "{On the solution of the radiative transfer equations for polarized radiation}",
      journal = {\solphys},
         year = 1985,
        month = jun,
       volume = {97},
        pages = {239-250},
          doi = {10.1007/BF00165988},
       adsurl = {https://ui.adsabs.harvard.edu/abs/1985SoPh...97..239L}
}

@ARTICLE{IPOLE_2018,
   author = {{Mo{\'s}cibrodzka}, M. and {Gammie}, C.~F.},
    title = "{IPOLE - semi-analytic scheme for relativistic polarized radiative transport}",
  journal = {\mnras},
archivePrefix = "arXiv",
   eprint = {1712.03057},
 primaryClass = "astro-ph.HE",
     year = 2018,
    month = mar,
   volume = 475,
    pages = {43-54},
      doi = {10.1093/mnras/stx3162},
   adsurl = {https://ui.adsabs.harvard.edu/abs/2018MNRAS.475...43M}
}

@ARTICLE{Wong_2022,
       author = {{Wong}, George N. and {Prather}, Ben S. and {Dhruv}, Vedant and {Ryan}, Benjamin R. and {Mo{\'s}cibrodzka}, Monika and {Chan}, Chi-kwan and {Joshi}, Abhishek V. and {Yarza}, Ricardo and {Ricarte}, Angelo and {Shiokawa}, Hotaka and {Dolence}, Joshua C. and {Noble}, Scott C. and {McKinney}, Jonathan C. and {Gammie}, Charles F.},
        title = "{PATOKA: Simulating Electromagnetic Observables of Black Hole Accretion}",
      journal = {\apjs},
         year = 2022,
        month = apr,
       volume = {259},
       number = {2},
          eid = {64},
        pages = {64},
          doi = {10.3847/1538-4365/ac582e},
archivePrefix = {arXiv},
       eprint = {2202.11721},
 primaryClass = {astro-ph.HE},
       adsurl = {https://ui.adsabs.harvard.edu/abs/2022ApJS..259...64W}
}

@ARTICLE{Qiu_2023,
       author = {{Qiu}, Richard and {Ricarte}, Angelo and {Narayan}, Ramesh and {Wong}, George N. and {Chael}, Andrew and {Palumbo}, Daniel},
        title = "{Using Machine Learning to link black hole accretion flows with spatially resolved polarimetric observables}",
      journal = {\mnras},
         year = 2023,
        month = apr,
       volume = {520},
       number = {4},
        pages = {4867-4888},
          doi = {10.1093/mnras/stad466},
archivePrefix = {arXiv},
       eprint = {2212.04852},
 primaryClass = {astro-ph.HE},
       adsurl = {https://ui.adsabs.harvard.edu/abs/2023MNRAS.520.4867Q}
}

@ARTICLE{PaperI,
   author = {{Event Horizon Telescope Collaboration} and {Akiyama}, K. and 
	{Alberdi}, A. and {Alef}, W. and {Asada}, K. and {Azulay}, R. and 
	{Baczko}, A.-K. and {Ball}, D. and {Balokovi{\'c}}, M. and {Barrett}, J. and et al.},
    title = "{First M87 Event Horizon Telescope Results. I. The Shadow of the Supermassive Black Hole}",
  journal = {\apjl},
archivePrefix = "arXiv",
   eprint = {1906.11238},
     year = "{2019a}",
    month = apr,
   volume = 875,
      eid = {L1},
    pages = {L1},
      doi = {10.3847/2041-8213/ab0ec7},
   adsurl = {https://ui.adsabs.harvard.edu/abs/2019ApJ...875L...1E}
}

@ARTICLE{PaperII,
   author = {{Event Horizon Telescope Collaboration} and {Akiyama}, K. and 
	{Alberdi}, A. and {Alef}, W. and {Asada}, K. and {Azulay}, R. and 
	{Baczko}, A.-K. and {Ball}, D. and {Balokovi{\'c}}, M. and {Barrett}, J. and et al.},
    title = "{First M87 Event Horizon Telescope Results. II. Array and Instrumentation}",
  journal = {\apjl},
archivePrefix = "arXiv",
   eprint = {1906.11239},
 primaryClass = "astro-ph.IM",
     year = "{2019b}",
    month = apr,
   volume = 875,
      eid = {L2},
    pages = {L2},
      doi = {10.3847/2041-8213/ab0c96},
   adsurl = {https://ui.adsabs.harvard.edu/abs/2019ApJ...875L...2E}
}

@ARTICLE{PaperIII,
   author = {{Event Horizon Telescope Collaboration} and {Akiyama}, K. and 
	{Alberdi}, A. and {Alef}, W. and {Asada}, K. and {Azulay}, R. and 
	{Baczko}, A.-K. and {Ball}, D. and {Balokovi{\'c}}, M. and {Barrett}, J. and et al.},
    title = "{First M87 Event Horizon Telescope Results. III. Data Processing and Calibration}",
  journal = {\apjl},
     year = "{2019c}",
    month = apr,
   volume = 875,
      eid = {L3},
    pages = {L3},
      doi = {10.3847/2041-8213/ab0c57},
   adsurl = {http://adsabs.harvard.edu/abs/2019ApJ...875L...3E}
}

@ARTICLE{PaperIV,
   author = {{Event Horizon Telescope Collaboration} and {Akiyama}, K. and 
	{Alberdi}, A. and {Alef}, W. and {Asada}, K. and {Azulay}, R. and 
	{Baczko}, A.-K. and {Ball}, D. and {Balokovi{\'c}}, M. and {Barrett}, J. and et al.},
    title = "{First M87 Event Horizon Telescope Results. IV. Imaging the Central Supermassive Black Hole}",
  journal = {\apjl},
     year = "{2019d}",
    month = apr,
   volume = 875,
      eid = {L4},
    pages = {L4},
      doi = {10.3847/2041-8213/ab0e85},
   adsurl = {https://ui.adsabs.harvard.edu/abs/2019ApJ...875L...4E}
}

@ARTICLE{PaperV,
   author = {{Event Horizon Telescope Collaboration} and {Akiyama}, K. and 
	{Alberdi}, A. and {Alef}, W. and {Asada}, K. and {Azulay}, R. and 
	{Baczko}, A.-K. and {Ball}, D. and {Balokovi{\'c}}, M. and {Barrett}, J. and et al.},
    title = "{First M87 Event Horizon Telescope Results. V. Physical Origin of the Asymmetric Ring}",
  journal = {\apjl},
archivePrefix = "arXiv",
   eprint = {1906.11242},
     year = "{2019e}",
    month = apr,
   volume = 875,
      eid = {L5},
    pages = {L5},
      doi = {10.3847/2041-8213/ab0f43},
   adsurl = {https://ui.adsabs.harvard.edu/abs/2019ApJ...875L...5E}
}

@ARTICLE{PaperVI,
   author = {{Event Horizon Telescope Collaboration} and {Akiyama}, K. and 
	{Alberdi}, A. and {Alef}, W. and {Asada}, K. and {Azulay}, R. and 
	{Baczko}, A.-K. and {Ball}, D. and {Balokovi{\'c}}, M. and {Barrett}, J. and et al.},
    title = "{First M87 Event Horizon Telescope Results. VI. The Shadow and Mass of the Central Black Hole}",
  journal = {\apjl},
archivePrefix = "arXiv",
   eprint = {1906.11243},
     year = "{2019f}",
    month = apr,
   volume = 875,
      eid = {L6},
    pages = {L6},
      doi = {10.3847/2041-8213/ab1141},
   adsurl = {https://ui.adsabs.harvard.edu/abs/2019ApJ...875L...6E}
}

@ARTICLE{Doeleman_2023,
       author = {{Doeleman}, Sheperd S. and {Barrett}, John and {Blackburn}, Lindy and {Bouman}, Katherine L. and {Broderick}, Avery E. and {Chaves}, Ryan and {Fish}, Vincent L. and {Fitzpatrick}, Garret and {Freeman}, Mark and {Fuentes}, Antonio and {G{\'o}mez}, Jos{\'e} L. and {Haworth}, Kari and {Houston}, Janice and {Issaoun}, Sara and {Johnson}, Michael D. and {Kettenis}, Mark and {Loinard}, Laurent and {Nagar}, Neil and {Narayanan}, Gopal and {Oppenheimer}, Aaron and {Palumbo}, Daniel C.~M. and {Patel}, Nimesh and {Pesce}, Dominic W. and {Raymond}, Alexander W. and {Roelofs}, Freek and {Srinivasan}, Ranjani and {Tiede}, Paul and {Weintroub}, Jonathan and {Wielgus}, Maciek},
        title = "{Reference Array and Design Consideration for the Next-Generation Event Horizon Telescope}",
      journal = {Galaxies},
         year = 2023,
        month = oct,
       volume = {11},
       number = {5},
          eid = {107},
        pages = {107},
          doi = {10.3390/galaxies11050107},
archivePrefix = {arXiv},
       eprint = {2306.08787},
 primaryClass = {astro-ph.IM},
       adsurl = {https://ui.adsabs.harvard.edu/abs/2023Galax..11..107D}
}

@ARTICLE{Johnson_2020,
       author = {{Johnson}, Michael D. and {Lupsasca}, Alexandru and {Strominger}, Andrew and {Wong}, George N. and {Hadar}, Shahar and {Kapec}, Daniel and {Narayan}, Ramesh and {Chael}, Andrew and {Gammie}, Charles F. and {Galison}, Peter and {Palumbo}, Daniel C.~M. and {Doeleman}, Sheperd S. and {Blackburn}, Lindy and {Wielgus}, Maciek and {Pesce}, Dominic W. and {Farah}, Joseph R. and {Moran}, James M.},
        title = "{Universal interferometric signatures of a black hole's photon ring}",
      journal = {Science Advances},
         year = 2020,
        month = mar,
       volume = {6},
       number = {12},
        pages = {eaaz1310},
          doi = {10.1126/sciadv.aaz1310},
archivePrefix = {arXiv},
       eprint = {1907.04329},
 primaryClass = {astro-ph.IM},
       adsurl = {https://ui.adsabs.harvard.edu/abs/2020SciA....6.1310J}
}

@ARTICLE{PaperVII,
       author = {{Event Horizon Telescope Collaboration} and {Akiyama}, Kazunori and {Algaba}, Juan Carlos and {Alberdi}, Antxon and {Alef}, Walter and {Anantua}, Richard and {Asada}, Keiichi and {Azulay}, Rebecca and {Baczko}, Anne-Kathrin and {Ball}, David and {Balokovi{\'c}}, Mislav and {Barrett}, John and {Benson}, Bradford A. and {Bintley}, Dan and {Blackburn}, Lindy and {Blundell}, Raymond and {Boland}, Wilfred and {Bouman}, Katherine L. and {Bower}, Geoffrey C. and {Boyce}, Hope and {Bremer}, Michael and {Brinkerink}, Christiaan D. and {Brissenden}, Roger and {Britzen}, Silke and {Broderick}, Avery E. and {Broguiere}, Dominique and {Bronzwaer}, Thomas and {Byun}, Do-Young and {Carlstrom}, John E. and {Chael}, Andrew and {Chan}, Chi-kwan and {Chatterjee}, Shami and {Chatterjee}, Koushik and {Chen}, Ming-Tang and {Chen}, Yongjun and {Chesler}, Paul M. and {Cho}, Ilje and {Christian}, Pierre and {Conway}, John E. and {Cordes}, James M. and {Crawford}, Thomas M. and {Crew}, Geoffrey B. and {Cruz-Osorio}, Alejandro and {Cui}, Yuzhu and {Davelaar}, Jordy and {De Laurentis}, Mariafelicia and {Deane}, Roger and {Dempsey}, Jessica and {Desvignes}, Gregory and {Dexter}, Jason and {Doeleman}, Sheperd S. and {Eatough}, Ralph P. and {Falcke}, Heino and {Farah}, Joseph and {Fish}, Vincent L. and {Fomalont}, Ed and {Ford}, H. Alyson and {Fraga-Encinas}, Raquel and {Freeman}, William T. and {Friberg}, Per and {Fromm}, Christian M. and {Fuentes}, Antonio and {Galison}, Peter and {Gammie}, Charles F. and {Garc{\'\i}a}, Roberto and {Gentaz}, Olivier and {Georgiev}, Boris and {Goddi}, Ciriaco and {Gold}, Roman and {G{\'o}mez}, Jos{\'e} L. and {G{\'o}mez-Ruiz}, Arturo I. and {Gu}, Minfeng and {Gurwell}, Mark and {Hada}, Kazuhiro and {Haggard}, Daryl and {Hecht}, Michael H. and {Hesper}, Ronald and {Ho}, Luis C. and {Ho}, Paul and {Honma}, Mareki and {Huang}, Chih-Wei L. and {Huang}, Lei and {Hughes}, David H. and {Ikeda}, Shiro and {Inoue}, Makoto and {Issaoun}, Sara and {James}, David J. and {Jannuzi}, Buell T. and {Janssen}, Michael and {Jeter}, Britton and {Jiang}, Wu and {Jimenez-Rosales}, Alejandra and {Johnson}, Michael D. and {Jorstad}, Svetlana and {Jung}, Taehyun and {Karami}, Mansour and {Karuppusamy}, Ramesh and {Kawashima}, Tomohisa and {Keating}, Garrett K. and {Kettenis}, Mark and {Kim}, Dong-Jin and {Kim}, Jae-Young and {Kim}, Jongsoo and {Kim}, Junhan and {Kino}, Motoki and {Koay}, Jun Yi and {Kofuji}, Yutaro and {Koch}, Patrick M. and {Koyama}, Shoko and {Kramer}, Michael and {Kramer}, Carsten and {Krichbaum}, Thomas P. and {Kuo}, Cheng-Yu and {Lauer}, Tod R. and {Lee}, Sang-Sung and {Levis}, Aviad and {Li}, Yan-Rong and {Li}, Zhiyuan and {Lindqvist}, Michael and {Lico}, Rocco and {Lindahl}, Greg and {Liu}, Jun and {Liu}, Kuo and {Liuzzo}, Elisabetta and {Lo}, Wen-Ping and {Lobanov}, Andrei P. and {Loinard}, Laurent and {Lonsdale}, Colin and {Lu}, Ru-Sen and {MacDonald}, Nicholas R. and {Mao}, Jirong and {Marchili}, Nicola and {Markoff}, Sera and {Marrone}, Daniel P. and {Marscher}, Alan P. and {Mart{\'\i}-Vidal}, Iv{\'a}n and {Matsushita}, Satoki and {Matthews}, Lynn D. and {Medeiros}, Lia and {Menten}, Karl M. and {Mizuno}, Izumi and {Mizuno}, Yosuke and {Moran}, James M. and {Moriyama}, Kotaro and {Moscibrodzka}, Monika and {M{\"u}ller}, Cornelia and {Musoke}, Gibwa and {Mej{\'\i}as}, Alejandro Mus and {Michalik}, Daniel and {Nadolski}, Andrew and {Nagai}, Hiroshi and {Nagar}, Neil M. and {Nakamura}, Masanori and {Narayan}, Ramesh and {Narayanan}, Gopal and {Natarajan}, Iniyan and {Nathanail}, Antonios and {Neilsen}, Joey and {Neri}, Roberto and {Ni}, Chunchong and {Noutsos}, Aristeidis and {Nowak}, Michael A. and {Okino}, Hiroki and {Olivares}, H{\'e}ctor and {Ortiz-Le{\'o}n}, Gisela N. and {Oyama}, Tomoaki and {{\"O}zel}, Feryal and {Palumbo}, Daniel C.~M. and {Park}, Jongho and {Patel}, Nimesh and {Pen}, Ue-Li and {Pesce}, Dominic W. and {Pi{\'e}tu}, Vincent and {Plambeck}, Richard and {PopStefanija}, Aleksandar and {Porth}, Oliver and {P{\"o}tzl}, Felix M. and {Prather}, Ben and {Preciado-L{\'o}pez}, Jorge A. and {Psaltis}, Dimitrios and {Pu}, Hung-Yi and {Ramakrishnan}, Venkatessh and {Rao}, Ramprasad and {Rawlings}, Mark G. and {Raymond}, Alexander W. and {Rezzolla}, Luciano and {Ricarte}, Angelo and {Ripperda}, Bart and {Roelofs}, Freek and {Rogers}, Alan and {Ros}, Eduardo and {Rose}, Mel and {Roshanineshat}, Arash and {Rottmann}, Helge and {Roy}, Alan L. and {Ruszczyk}, Chet and {Rygl}, Kazi L.~J. and {S{\'a}nchez}, Salvador and {S{\'a}nchez-Arguelles}, David and {Sasada}, Mahito and {Savolainen}, Tuomas and {Schloerb}, F. Peter and {Schuster}, Karl-Friedrich and {Shao}, Lijing and {Shen}, Zhiqiang and {Small}, Des and {Sohn}, Bong Won and {SooHoo}, Jason and {Sun}, He and {Tazaki}, Fumie and {Tetarenko}, Alexandra J. and {Tiede}, Paul and {Tilanus}, Remo P.~J. and {Titus}, Michael and {Toma}, Kenji and {Torne}, Pablo and {Trent}, Tyler and {Traianou}, Efthalia and {Trippe}, Sascha and {van Bemmel}, Ilse and {van Langevelde}, Huib Jan and {van Rossum}, Daniel R. and {Wagner}, Jan and {Ward-Thompson}, Derek and {Wardle}, John and {Weintroub}, Jonathan and {Wex}, Norbert and {Wharton}, Robert and {Wielgus}, Maciek and {Wong}, George N. and {Wu}, Qingwen and {Yoon}, Doosoo and {Young}, Andr{\'e} and {Young}, Ken and {Younsi}, Ziri and {Yuan}, Feng and {Yuan}, Ye-Fei and {Zensus}, J. Anton and {Zhao}, Guang-Yao and {Zhao}, Shan-Shan},
        title = "{First M87 Event Horizon Telescope Results. VII. Polarization of the Ring}",
      journal = {\apjl},
         year = 2021,
        month = mar,
       volume = {910},
       number = {1},
          eid = {L12},
        pages = {L12},
          doi = {10.3847/2041-8213/abe71d},
archivePrefix = {arXiv},
       eprint = {2105.01169},
 primaryClass = {astro-ph.HE},
       adsurl = {https://ui.adsabs.harvard.edu/abs/2021ApJ...910L..12E}
}

@ARTICLE{PaperVIII,
       author = {{Event Horizon Telescope Collaboration} and {Akiyama}, Kazunori and {Algaba}, Juan Carlos and {Alberdi}, Antxon and {Alef}, Walter and {Anantua}, Richard and {Asada}, Keiichi and {Azulay}, Rebecca and {Baczko}, Anne-Kathrin and {Ball}, David and {Balokovi{\'c}}, Mislav and {Barrett}, John and {Benson}, Bradford A. and {Bintley}, Dan and {Blackburn}, Lindy and {Blundell}, Raymond and {Boland}, Wilfred and {Bouman}, Katherine L. and {Bower}, Geoffrey C. and {Boyce}, Hope and {Bremer}, Michael and {Brinkerink}, Christiaan D. and {Brissenden}, Roger and {Britzen}, Silke and {Broderick}, Avery E. and {Broguiere}, Dominique and {Bronzwaer}, Thomas and {Byun}, Do-Young and {Carlstrom}, John E. and {Chael}, Andrew and {Chan}, Chi-kwan and {Chatterjee}, Shami and {Chatterjee}, Koushik and {Chen}, Ming-Tang and {Chen}, Yongjun and {Chesler}, Paul M. and {Cho}, Ilje and {Christian}, Pierre and {Conway}, John E. and {Cordes}, James M. and {Crawford}, Thomas M. and {Crew}, Geoffrey B. and {Cruz-Osorio}, Alejandro and {Cui}, Yuzhu and {Davelaar}, Jordy and {De Laurentis}, Mariafelicia and {Deane}, Roger and {Dempsey}, Jessica and {Desvignes}, Gregory and {Dexter}, Jason and {Doeleman}, Sheperd S. and {Eatough}, Ralph P. and {Falcke}, Heino and {Farah}, Joseph and {Fish}, Vincent L. and {Fomalont}, Ed and {Ford}, H. Alyson and {Fraga-Encinas}, Raquel and {Friberg}, Per and {Fromm}, Christian M. and {Fuentes}, Antonio and {Galison}, Peter and {Gammie}, Charles F. and {Garc{\'\i}a}, Roberto and {Gelles}, Zachary and {Gentaz}, Olivier and {Georgiev}, Boris and {Goddi}, Ciriaco and {Gold}, Roman and {G{\'o}mez}, Jos{\'e} L. and {G{\'o}mez-Ruiz}, Arturo I. and {Gu}, Minfeng and {Gurwell}, Mark and {Hada}, Kazuhiro and {Haggard}, Daryl and {Hecht}, Michael H. and {Hesper}, Ronald and {Himwich}, Elizabeth and {Ho}, Luis C. and {Ho}, Paul and {Honma}, Mareki and {Huang}, Chih-Wei L. and {Huang}, Lei and {Hughes}, David H. and {Ikeda}, Shiro and {Inoue}, Makoto and {Issaoun}, Sara and {James}, David J. and {Jannuzi}, Buell T. and {Janssen}, Michael and {Jeter}, Britton and {Jiang}, Wu and {Jimenez-Rosales}, Alejandra and {Johnson}, Michael D. and {Jorstad}, Svetlana and {Jung}, Taehyun and {Karami}, Mansour and {Karuppusamy}, Ramesh and {Kawashima}, Tomohisa and {Keating}, Garrett K. and {Kettenis}, Mark and {Kim}, Dong-Jin and {Kim}, Jae-Young and {Kim}, Jongsoo and {Kim}, Junhan and {Kino}, Motoki and {Koay}, Jun Yi and {Kofuji}, Yutaro and {Koch}, Patrick M. and {Koyama}, Shoko and {Kramer}, Michael and {Kramer}, Carsten and {Krichbaum}, Thomas P. and {Kuo}, Cheng-Yu and {Lauer}, Tod R. and {Lee}, Sang-Sung and {Levis}, Aviad and {Li}, Yan-Rong and {Li}, Zhiyuan and {Lindqvist}, Michael and {Lico}, Rocco and {Lindahl}, Greg and {Liu}, Jun and {Liu}, Kuo and {Liuzzo}, Elisabetta and {Lo}, Wen-Ping and {Lobanov}, Andrei P. and {Loinard}, Laurent and {Lonsdale}, Colin and {Lu}, Ru-Sen and {MacDonald}, Nicholas R. and {Mao}, Jirong and {Marchili}, Nicola and {Markoff}, Sera and {Marrone}, Daniel P. and {Marscher}, Alan P. and {Mart{\'\i}-Vidal}, Iv{\'a}n and {Matsushita}, Satoki and {Matthews}, Lynn D. and {Medeiros}, Lia and {Menten}, Karl M. and {Mizuno}, Izumi and {Mizuno}, Yosuke and {Moran}, James M. and {Moriyama}, Kotaro and {Moscibrodzka}, Monika and {M{\"u}ller}, Cornelia and {Musoke}, Gibwa and {Mus Mej{\'\i}as}, Alejandro and {Michalik}, Daniel and {Nadolski}, Andrew and {Nagai}, Hiroshi and {Nagar}, Neil M. and {Nakamura}, Masanori and {Narayan}, Ramesh and {Narayanan}, Gopal and {Natarajan}, Iniyan and {Nathanail}, Antonios and {Neilsen}, Joey and {Neri}, Roberto and {Ni}, Chunchong and {Noutsos}, Aristeidis and {Nowak}, Michael A. and {Okino}, Hiroki and {Olivares}, H{\'e}ctor and {Ortiz-Le{\'o}n}, Gisela N. and {Oyama}, Tomoaki and {{\"O}zel}, Feryal and {Palumbo}, Daniel C.~M. and {Park}, Jongho and {Patel}, Nimesh and {Pen}, Ue-Li and {Pesce}, Dominic W. and {Pi{\'e}tu}, Vincent and {Plambeck}, Richard and {PopStefanija}, Aleksandar and {Porth}, Oliver and {P{\"o}tzl}, Felix M. and {Prather}, Ben and {Preciado-L{\'o}pez}, Jorge A. and {Psaltis}, Dimitrios and {Pu}, Hung-Yi and {Ramakrishnan}, Venkatessh and {Rao}, Ramprasad and {Rawlings}, Mark G. and {Raymond}, Alexander W. and {Rezzolla}, Luciano and {Ricarte}, Angelo and {Ripperda}, Bart and {Roelofs}, Freek and {Rogers}, Alan and {Ros}, Eduardo and {Rose}, Mel and {Roshanineshat}, Arash and {Rottmann}, Helge and {Roy}, Alan L. and {Ruszczyk}, Chet and {Rygl}, Kazi L.~J. and {S{\'a}nchez}, Salvador and {S{\'a}nchez-Arguelles}, David and {Sasada}, Mahito and {Savolainen}, Tuomas and {Schloerb}, F. Peter and {Schuster}, Karl-Friedrich and {Shao}, Lijing and {Shen}, Zhiqiang and {Small}, Des and {Sohn}, Bong Won and {SooHoo}, Jason and {Sun}, He and {Tazaki}, Fumie and {Tetarenko}, Alexandra J. and {Tiede}, Paul and {Tilanus}, Remo P.~J. and {Titus}, Michael and {Toma}, Kenji and {Torne}, Pablo and {Trent}, Tyler and {Traianou}, Efthalia and {Trippe}, Sascha and {van Bemmel}, Ilse and {van Langevelde}, Huib Jan and {van Rossum}, Daniel R. and {Wagner}, Jan and {Ward-Thompson}, Derek and {Wardle}, John and {Weintroub}, Jonathan and {Wex}, Norbert and {Wharton}, Robert and {Wielgus}, Maciek and {Wong}, George N. and {Wu}, Qingwen and {Yoon}, Doosoo and {Young}, Andr{\'e} and {Young}, Ken and {Younsi}, Ziri and {Yuan}, Feng and {Yuan}, Ye-Fei and {Zensus}, J. Anton and {Zhao}, Guang-Yao and {Zhao}, Shan-Shan},
        title = "{First M87 Event Horizon Telescope Results. VIII. Magnetic Field Structure near The Event Horizon}",
      journal = {\apjl},
         year = 2021,
        month = mar,
       volume = {910},
       number = {1},
          eid = {L13},
        pages = {L13},
          doi = {10.3847/2041-8213/abe4de},
archivePrefix = {arXiv},
       eprint = {2105.01173},
 primaryClass = {astro-ph.HE},
       adsurl = {https://ui.adsabs.harvard.edu/abs/2021ApJ...910L..13E}
}

@ARTICLE{Pu_2018,
       author = {{Pu}, Hung-Yi and {Broderick}, Avery E.},
        title = "{Probing the Innermost Accretion Flow Geometry of Sgr A* with Event Horizon Telescope}",
      journal = {\apj},
         year = 2018,
        month = aug,
       volume = {863},
       number = {2},
          eid = {148},
        pages = {148},
          doi = {10.3847/1538-4357/aad086},
archivePrefix = {arXiv},
       eprint = {1807.01817},
 primaryClass = {astro-ph.HE},
       adsurl = {https://ui.adsabs.harvard.edu/abs/2018ApJ...863..148P}
}

@INPROCEEDINGS{Bardeen_1973,
       author = {{Bardeen}, J.~M.},
        title = "{Timelike and null geodesics in the Kerr metric.}",
    booktitle = {Black Holes (Les Astres Occlus)},
         year = 1973,
        month = jan,
        pages = {215-239},
       adsurl = {https://ui.adsabs.harvard.edu/abs/1973blho.conf..215B}
}

@ARTICLE{Johannsen_2010,
       author = {{Johannsen}, Tim and {Psaltis}, Dimitrios},
        title = "{Testing the No-hair Theorem with Observations in the Electromagnetic Spectrum. II. Black Hole Images}",
      journal = {\apj},
         year = 2010,
        month = jul,
       volume = {718},
       number = {1},
        pages = {446-454},
          doi = {10.1088/0004-637X/718/1/446},
archivePrefix = {arXiv},
       eprint = {1005.1931},
 primaryClass = {astro-ph.HE},
       adsurl = {https://ui.adsabs.harvard.edu/abs/2010ApJ...718..446J}
}

@ARTICLE{Gelles_2021,
       author = {{Gelles}, Zachary and {Himwich}, Elizabeth and {Palumbo}, Daniel C.~M. and {Johnson}, Michael D.},
        title = "{Polarized Image of Equatorial Emission in the Kerr Geometry}",
      journal = {arXiv e-prints},
         year = 2021,
        month = may,
          eid = {arXiv:2105.09440},
        pages = {arXiv:2105.09440},
archivePrefix = {arXiv},
       eprint = {2105.09440},
 primaryClass = {gr-qc},
       adsurl = {https://ui.adsabs.harvard.edu/abs/2021arXiv210509440G}
}

@ARTICLE{Mosci_2016,
   author = {{Mo{\'s}cibrodzka}, M. and {Falcke}, H. and {Shiokawa}, H.},
    title = "{General relativistic magnetohydrodynamical simulations of the jet in M 87}",
  journal = {\aap},
archivePrefix = "arXiv",
   eprint = {1510.07243},
 primaryClass = "astro-ph.HE",
     year = 2016,
    month = feb,
   volume = 586,
      eid = {A38},
    pages = {A38},
      doi = {10.1051/0004-6361/201526630},
   adsurl = {http://adsabs.harvard.edu/abs/2016A%26A...586A..38M}
}

@ARTICLE{Narayan_2022,
       author = {{Narayan}, Ramesh and {Chael}, Andrew and {Chatterjee}, Koushik and {Ricarte}, Angelo and {Curd}, Brandon},
        title = "{Jets in magnetically arrested hot accretion flows: geometry, power, and black hole spin-down}",
      journal = {\mnras},
         year = 2022,
        month = apr,
       volume = {511},
       number = {3},
        pages = {3795-3813},
          doi = {10.1093/mnras/stac285},
archivePrefix = {arXiv},
       eprint = {2108.12380},
 primaryClass = {astro-ph.HE},
       adsurl = {https://ui.adsabs.harvard.edu/abs/2022MNRAS.511.3795N}
}

@BOOK{R&L,
       author = {{Rybicki}, George B. and {Lightman}, Alan P.},
        title = "{Radiative processes in astrophysics}",
         year = 1979,
       adsurl = {https://ui.adsabs.harvard.edu/abs/1979rpa..book.....R}
}

@ARTICLE{Wong_2025,
       author = {{Wong}, George N. and {Chael}, Andrew and {Lupsasca}, Alexandru and {Quataert}, Eliot},
        title = "{Black Hole Polarimetry II: The Connection Between Spin and Polarization}",
      journal = {arXiv e-prints},
         year = 2025,
        month = sep,
          eid = {arXiv:2509.22639},
        pages = {arXiv:2509.22639},
          doi = {10.48550/arXiv.2509.22639},
archivePrefix = {arXiv},
       eprint = {2509.22639},
 primaryClass = {astro-ph.HE},
       adsurl = {https://ui.adsabs.harvard.edu/abs/2025arXiv250922639W}
}

\end{document}